\documentclass[preprint, 12pt]{revtex4-2}
\usepackage[utf8]{inputenc}
\usepackage{geometry}
\usepackage{amsmath}
\usepackage{graphicx}
\usepackage{caption}
\usepackage{subcaption}
\usepackage{color}
\usepackage[T1]{fontenc}
\begin{document}

\title{Engineering the Radiative Dynamics of Thermalized Excitons with Metal Interfaces}

\author{Grace H. Chen$^1$}
\author{David Z. Li$^2$}
\author{Amy Butcher$^1$}
\author{Alexander A. High$^{1, 3}$}
\author{Darrick E. Chang$^{2, 4}$}

\affiliation{$^1$Pritzker School of Molecular Engineering,
University of Chicago, Chicago, Illinois 60637, United States}
\affiliation{$^2$ICFO-Institut de Ciencies Fotoniques, The Barcelona Institute of Science and Technology, 08860 Castelldefels (Barcelona), Spain}
\affiliation{$^3$Center for Molecular Engineering and Materials Science
Division, Argonne National Laboratory, Lemont, Illinois 60439,
United States}
\affiliation{$^4$ICREA-Institució Catalana de Recerca i Estudis Avançats, 08015 Barcleona, Spain}

\begin{abstract}
    As a platform for optoelectronic devices based on exciton dynamics, monolayer transition metal dichalcogenides (TMDCs) are often placed near metal interfaces or inside planar cavities. While the radiative properties of point dipoles at metal interfaces has been studied extensively, those of excitons, which are delocalized and exhibit a temperature-dependent momentum distribution, lack a thorough treatment. Here, we analyze the emission properties of excitons in TMDCs near planar metal interfaces and explore their dependence on exciton center-of-mass momentum, transition dipole orientation, and temperature. Defining a characteristic energy scale $k_B T_c = (\hbar k)^2/2m$~($k$ being the radiative wavevector and $m$ the exciton mass), we find that at temperatures $T\gg T_c$ and low densities where the momentum distribution can be characterized by Maxwell-Boltzmann statistics, the modified emission rates~(normalized to free space) behave similarly to point dipoles at temperatures $T\gg T_c$. This similarity in behavior arises due to the broad nature of wavevector components making up the exciton and point dipole emission. On the other hand, the narrow momentum distribution of excitons for $T<T_c$ can result in significantly different emission behavior as compared to point dipoles. These differences can be further amplified by considering excitons with a Bose Einstein distribution at high phase space densities. We find suppression or enhancement of emission relative to the point dipole case by several orders of magnitude. These insights can help optimize the performance of optoelectronic devices that incorporate 2D semiconductors near metal electrodes and can inform future studies of exciton radiative dynamics at low temperatures. Additionally, these studies show that nanoscale optical cavities are a viable pathway to generating long-lifetime exciton states in TMDCs.
\end{abstract}

\maketitle
\newpage 
\section{Introduction}


Excitons in two-dimensional quantum wells are widely utilized in optoelectronic technologies [1,2] and in studies of bosonic superfluidity and condensation [3-5]. In particular, excitons in monolayer transition metal dichalcogenides (TMDCs) have attracted significant recent interest due to their large binding energies, which enable condensed and superfluid phases to emerge at high temperatures, and their strong oscillator strength, which provides a strong, coherent optical response to near-resonant light [1, 4]. These properties, combined with the fundamentally extended nature of the center-of-mass wave function, can give rise to interesting functionalities, such as tunable, atomically thin reflecting elements [6-9]. \par

Engineering exciton lifetimes with nanostructures can advance numerous applications such as integration with electrically tunable interlayer exciton systems or studies of quantum phases [3, 5, 10]. Certainly, the ability to modify emission of point-like emitters has been well-studied, and forms the basis of important applications such as quantum information processing [11-14] and single-molecule detection [15, 16]. However, the principles and intuition developed for point emitters do not necessarily apply to excitons, which, in contrast to point dipoles, are fundamentally delocalized excitations with a temperature-dependent momentum distribution. Typically, long exciton lifetimes are achieved in quantum well heterostructures, in which the electron and hole wavefunctions are physically separated, which suppresses radiative recombination and allows for the creation of a thermalized, high density gas [3]. These interlayer systems have the same selection rules as monolayer TMDCs, with circularly polarized, in-plane transitions [17]. Recently, these thermalized, high density exciton ensembles have shown experimental signatures of Bose-condensed phases with a non-classical occupation of low momentum states [4, 5], significantly impacting both the radiative coupling of excitons to cavity structures and the non-radiative dynamics of excitons in proximity to metal films. This non-classical occupation can also yield different mechanisms for emission suppression or enhancement [7]. An alternative pathway to engineer the radiative dynamics is through local engineering of the optical environment [18]. Importantly, these TMDC heterostructures and other quantum well systems are frequently placed in close proximity to metal films, which are utilized as gate electrodes or as plasmonic substrates [3, 19]. For point dipoles, this proximity creates non-radiative quenching that exponentially increases with proximity to the metal, limiting optical applications that rely on high radiative efficiency [20]. Alternatively, the temperature dependent wavevector distribution of excitons creates a mechanism to avoid detrimental quenching processes.

Here, we theoretically investigate the impact of two planar geometries, a single planar metal interface and a planar metal cavity, on the radiative dynamics of excitons, taking into account the thermal distribution of exciton momentum. 
We investigate the dependence of the emission rates on the orientation (in- or out-of-plane) of the dipole transition, the distance between the TMDC and the interface(s), the center-of-mass momentum ${\bf Q}$ of the exciton, and the momentum distribution, assuming that it can be described as thermal. For these thermal systems, we find that the emission rate at high temperatures, when normalized by the free-space emission at the same temperature, has a functional form that closely approximates the emission expected of a point dipole emitter at all distances. In contrast, the divergent emission rate found at small distances for point dipoles (associated with Ohmic dissipation in the metal) is largely suppressed at low temperatures ($< 10$ K for single interface and planar cavities). Finally, we find that for a fixed density ($n \sim 10^{12}$ cm$^{-2}$), the calculation of radiative decay rates using Bose-Einstein statistical distributions can enhance or suppress emission by orders of magnitude at typical experimental temperatures ($\sim 5$ K). \par

In section \ref{sec:theory}, we introduce the theoretical formalism to calculate modified spontaneous emission rates in terms of the total field seen by classical radiating point or extended ``planar'' dipoles at their own locations. In the latter case, which is valid to describe extended excitons, the emission rate is calculated as a function of center-of-mass momentum ${\bf Q}$. In section \ref{sec:dipres} we apply this general formalism to derive specific rates in the vicinity of a single interface and in a planar cavity. In section \ref{sec:temp}, we introduce our model for the temperature dependent emission rate of an extended exciton, assuming that the center of mass momentum is characterized by a Bose-Einstein distribution. In sections~\ref{sec:SI} and~\ref{sec:CAV}, we present numerical results for the temperature and distance-dependent emission rates for a single interface and cavity, specifically considering a silver structure. For a distribution at low phase space density~(i.e. a Maxwell-Boltzmann distribution or the limit of a single exciton), we show that the emission rate~(normalized to that of free space at the same temperature) approaches that of a point dipole at high temperatures $T\gg T_c$, where $k_B T_c=(\hbar k)^2/2m$~($k=2\pi/\lambda$ being the radiative wavevector and $m$ the exciton mass). In the opposite limit of $T \lesssim T_c$, the behavior can significantly differ from a point dipole. This difference is clearest at close distances to the interface ($d\lesssim\lambda/2\pi$), as the narrow momentum distribution highly suppresses non-radiative emission, and at large distances when this narrow distribution is able to resolve the resonances of the cavity structure. Although $T_c\sim 30$ mK can be quite low for experimental realization, at high phase space densities, the narrow momentum distribution arising from quasi-condensation allows for such changes in emission to be observed even at relatively high temperatures ($\sim 5$K). 
In section~\ref{sec:conclusion} we provide a brief conclusion and outlook of our work.

\section{Theoretical formalism \label{sec:theory}}

\begin{figure}[h!]
    \centering
    \includegraphics[width = \textwidth]{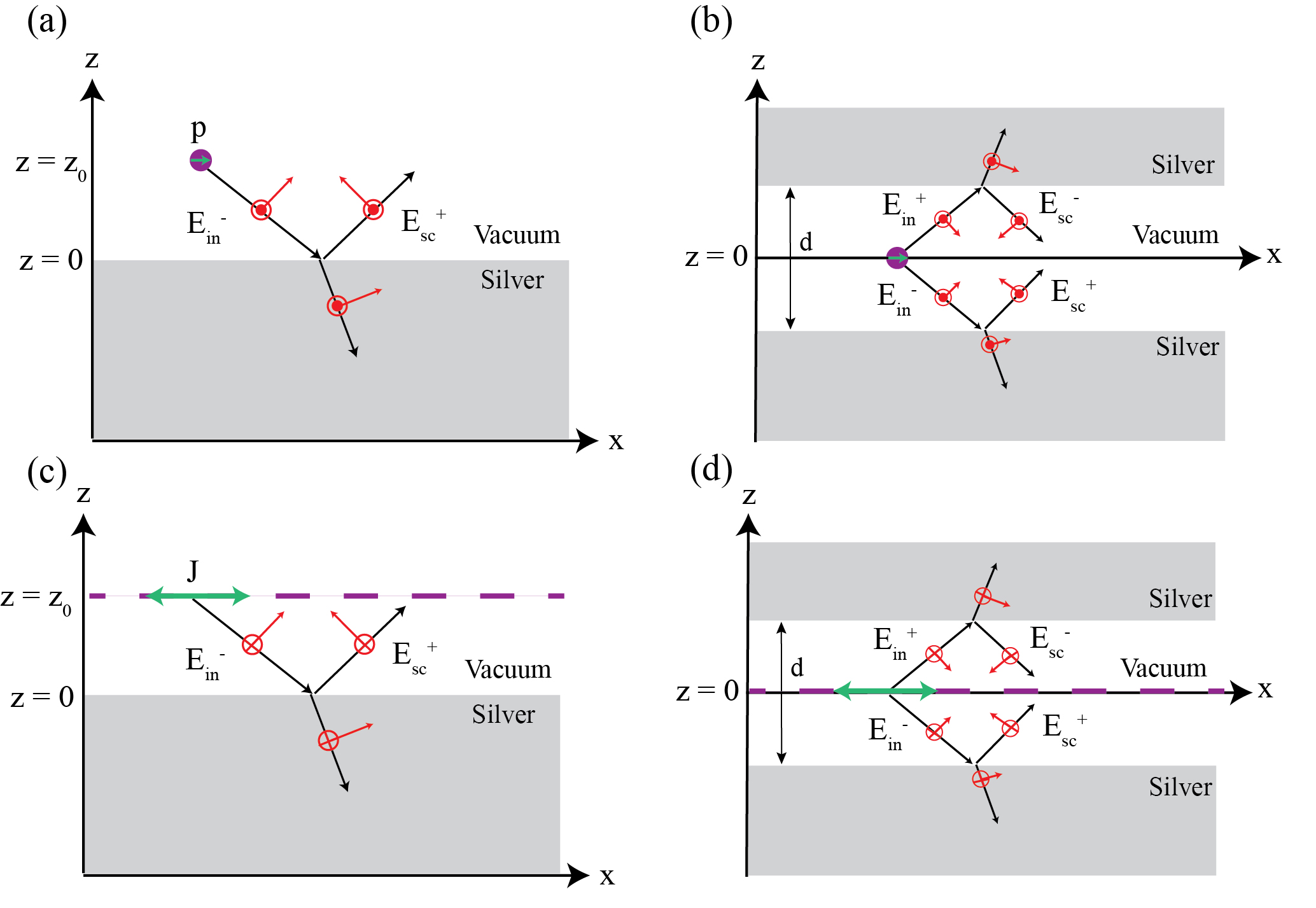}
    \caption{(a) Point dipole with in-plane polarization, at a distance $z_0$ above a silver interface. (b) Point dipole with in-plane polarization, located symmetrically between a silver cavity with spacing $d$ between mirrors. (c) Planar dipole with in-plane polarization at a distance $z_0$ above a silver interface. (d) Planar dipole with in-plane polarization, located symmetrically between a silver cavity. The emitted light in all cases can be decomposed into different wavevectors whose directions are indicated by the black arrows~(showing both incident and scattered fields from the surface), while the red arrows and red circles/crosses denote the polarizations associated with s- and p-polarized light. Analogously, we also consider the case of dipoles with out-of-plane polarization~(not illustrated).}
    \label{fig:schem}
\end{figure}

In this section, we briefly review the theory by which modified spontaneous emission rates can be calculated. We first begin with the better known case of a point dipole, before presenting the ``planar'' dipole used to model an extended exciton. We present the geometry for each structure in the point and planar dipole cases in figure \ref{fig:schem}.

\captionsetup[figure]{justification=raggedright, labelfont=bf}

\subsection{Point-dipole Emission Rates}
The point dipole case is particularly prevalent in quantum optics, accounting for the modification of spontaneous emission of individual atoms, molecules or other point-like quantum emitters in the vicinity of planar interfaces [20, 21], planar cavities [22], or other resonant or confining structures for light [23-25]. \par
We consider a two-level system with an electric dipole allowed transition of frequency $\omega$ and corresponding free-space wavelength $\lambda$. We assume that its immediate vicinity is characterized by a relative permittivity $\epsilon_1$, but allow for the possibility of other dielectric media nearby, which might modify its emission rate. In the weak-coupling regime (to be defined shortly), the modified emission rate, $\Gamma$, normalized by the emission rate $\Gamma_0$ in an infinite medium of permittivity $\epsilon_1$, is given by [20, 26, 27]:
\begin{align}
    \frac{\Gamma}{\Gamma_0} &= \frac{\textrm{Im}\left[ \textbf{p}^* \cdot \textbf{E}(\boldsymbol{r_0}) \right]}{\textrm{Im}\left[ \textbf{p}^* \cdot \textbf{E}_{free}(\boldsymbol{r_0}) \right]}. \label{eq:dip}
\end{align}
Here, $\textbf{p}e^{-i\omega t}$ is a classical oscillating point dipole representing the two-level system,  $\textbf{E}(\boldsymbol{r_0})e^{-i\omega t}$ is the total field produced by the dipole in the geometry of interest at its own location $\boldsymbol{r_0} = (0, 0, z_0)$, and $\textbf{E}_{free}(\boldsymbol{r_0})$ is the field of the dipole in a uniform medium with a relative permittivity of $\epsilon_1$.
The validity of this weak-coupling approach requires that the response of the electromagnetic environment itself does not appreciably vary over $\Gamma$ itself. Otherwise, non-exponential or strong coupling dynamics like vacuum Rabi oscillations can occur [28]. \par
While equation (\ref{eq:dip}) is general, we now describe an efficient method to calculate the total field $\textbf{E}$ for planar geometries. We outline the calculations for a point dipole with arbitrary orientation, and for convenience define the corresponding current density $\textbf{J}(\boldsymbol{r}) = -i \omega \delta(\boldsymbol{r} - \boldsymbol{r_0}) \textbf{p}$. We first define the in-plane wavevector $\boldsymbol{k_\parallel} = (k_x, k_y)$, and generalizing for later discussions, the dispersion relation in medium $i$ of $k_\parallel^2 + k_{z,i}^2 = \epsilon_i \left( \frac{\omega}{c}\right)^2\equiv k_i^2$. The free field in an infinite medium with permittivity $\epsilon$ can be expressed as:
\begin{align}
    \textbf{E}_{free}(\boldsymbol{r}) &= i \omega \mu_0 \int d\boldsymbol{r'} \textbf{G}_{free}(\boldsymbol{r}, \boldsymbol{r'}) \cdot \textbf{J}(\boldsymbol{r'}) \label{eq:efieldgreen}
\end{align}
For planar geometries, it is useful to express the dyadic Green's function in a plane wave representation:
\begin{align}
    \textbf{G}_{free}(\boldsymbol{r}, \boldsymbol{r'}) &= \frac{i}{8 \pi^2} \int d^2 \boldsymbol{k_\parallel}  \textbf{M}(\boldsymbol{k_\parallel}) e^{i(\boldsymbol{k_\parallel} \cdot (\boldsymbol{\rho} - \boldsymbol{\rho'})+ k_z |z-z'|)} \label{eq:greens}
\end{align}
and:
\begin{align}
    \textbf{M}(\boldsymbol{k_\parallel}) &= \left( \frac{1}{k_z k^2}\right)
    \begin{pmatrix}
    k^2-k_x^2 & -k_x k_y & \mp k_x k_z \\
    -k_x k_y & k^2-k_y^2 & \mp k_y k_z \\
    \mp k_x k_z & \mp k_y k_z & k^2-k_z^2
    \end{pmatrix} \label{eq:m}
\end{align}
(Note that the upper sign is for $z>z_0$ and the lower sign for $z<z_0$) [27]. For the point dipole, this expression for the field simplifies to:
\begin{align}
    \textbf{E}_{free} (\boldsymbol{r}) &= \frac{1}{4 \pi^2} \int d^2\boldsymbol{k_\parallel} \textbf{ E}(\boldsymbol{k_\parallel}) e^{i(\boldsymbol{k_\parallel} \cdot \boldsymbol{\rho} + k_{z} |z-z_0|)} \label{eq:dipfield}
\end{align}
where
\begin{align}
    \textbf{E}(\boldsymbol{k_\parallel}) &= \frac{i \omega^2 \mu_0}{2} \textbf{M}(\boldsymbol{k_\parallel}) \cdot  \textbf{p} \label{eq:amplitude}
\end{align}
gives the polarization and amplitude of each plane wave component. \par

To calculate the total field, we write it as a sum of free and scattered contributions, $\textbf{E}(\boldsymbol{r})=\textbf{E}_{free}(\boldsymbol{r})+\textbf{E}_{sc}(\boldsymbol{r})$ and solve for the latter. For concreteness, we will focus on two cases, consisting of a single dielectric interface (figure \ref{fig:schem}(a)) and a symmetric cavity configuration (figure \ref{fig:schem}(b)). For the single interface, we consider that the point dipole is situated at a position $\boldsymbol{r_0}$~(with $z_0>0$) in a medium with permittivity $\epsilon_1$, and calculate the reflected field from an interface with a material with permittivity $\epsilon_2$, with the boundary between the two media situated at $z=0$. For an arbitrarily polarized point-dipole $\textbf{p}$, it is useful to decompose the fields into s- and p- polarizations to calculate the reflected~(i.e. scattered) fields. We decompose the matrix $\textbf{M}(\boldsymbol{k_\parallel}) = \textbf{M}^{(s)}(\boldsymbol{k_\parallel})+ \textbf{M}^{(p)}(\boldsymbol{k_\parallel})$ where:
\begin{align}
    \textbf{M}^{(s)}(\boldsymbol{k_\parallel}) &= \left( \frac{1}{k_{z,1} k_\parallel^2}\right)
    \begin{pmatrix}
    k_y^2 & -k_x k_y & 0 \\
    -k_x k_y & k_x^2 & 0 \\
    0 & 0 & 0
    \end{pmatrix} \label{eq:ms} \\ 
    \textbf{M}^{(p)}(\boldsymbol{k_\parallel}) &= \left( \frac{1}{k_{1}^2 k_\parallel^2}\right)
    \begin{pmatrix}
    k_x^2 k_{z,1} & k_x k_y k_{z,1} & \mp k_x k_\parallel^2 \\
    k_x k_y k_{z,1} & k_y^2 k_{z,1} & \mp k_y k_\parallel^2 \\
    \mp k_x k_\parallel^2 & \mp k_y k_\parallel^2 & \frac{k_\parallel^4}{k_{z,1}}
    \end{pmatrix} \label{eq:mp}
\end{align}
gives the decomposed matrices for the incident fields and
\begin{align}
    \textbf{M}^{(s)}_R(\boldsymbol{k_\parallel}) &= \left( \frac{1}{k_{z,1} k_\parallel^2}\right)
    \begin{pmatrix}
    k_y^2 & -k_x k_y & 0 \\
    -k_x k_y & k_x^2 & 0 \\
    0 & 0 & 0
    \end{pmatrix} \label{eq:msr} \\
    \textbf{M}^{(p, \pm)}_R(\boldsymbol{k_\parallel}) &= \left(- \frac{1}{k_{1}^2 k_\parallel^2}\right)
    \begin{pmatrix}
    k_x^2 k_{z,1} & k_x k_y k_{z,1} & \pm k_x k_\parallel^2 \\
    k_x k_y k_{z,1} & k_y^2 k_{z,1} & \pm k_y k_\parallel^2 \\
    \mp k_x k_\parallel^2 & \mp k_y k_\parallel^2 & \mp \frac{k_\parallel^4}{k_{z,1}}
    \end{pmatrix} \label{eq:mpr}
\end{align}
gives the decomposed matrices for the reflected fields [27]. Note that in the p-polarized reflection matrix, the $\pm$ superscript denotes the propagation direction of the reflected field. For the geometry in figure~\ref{fig:schem}(a), the reflected field propagates upward and corresponds to the choice $+$. \par

For the single interface, the reflected field at the location of the point dipole is:
\begin{align}
    \textbf{E}_{sc} (\boldsymbol{r_0}) &= \frac{1}{4 \pi^2} \int d^2 \boldsymbol{k_\parallel} \boldsymbol{E_{sc}} (\boldsymbol{k_\parallel}) e^{2 i k_{z,1} z_0}
\end{align}
where
\begin{align}
    \boldsymbol{E_{sc}} (\boldsymbol{k_\parallel}) &= \frac{i \omega^2 \mu_0}{2} \left[r_s(\boldsymbol{k_\parallel}) \textbf{M}_R^{(s)}(\boldsymbol{k_\parallel}) + r_p(\boldsymbol{k_\parallel}) \textbf{M}_R^{(p, +)} (\boldsymbol{k_\parallel}) \right] \cdot  \textbf{p}.
\end{align}
The reflection coefficients of a plane wave from a planar interface $r_{s,p}(\boldsymbol{k_\parallel})$ are given by [27, 29]:
\begin{align}
    r_s(\boldsymbol{k_\parallel}) = \frac{k_{z,1}-k_{z,2}}{k_{z,1}+k_{z,2}}  \qquad
    r_p (\boldsymbol{k_\parallel}) = \frac{\epsilon_2 k_{z,1} - \epsilon_1 k_{z,2}}{\epsilon_2 k_{z,1} + \epsilon_1 k_{z,2}}. \label{eq:rsp}
\end{align}
One can now substitute the total~(free plus scattered) field into equation (\ref{eq:dip}) to calculate the modified emission rate at a single interface. We consider two specific orientations of the point dipole: in-plane ($\textbf{p} = (p_x, p_y, 0)$) and out-of-plane ($\textbf{p} = (0, 0, p)$) to the interface. The total emission rate for an in-plane dipole is:
\begin{align}
    \frac{\Gamma (z_0)}{\Gamma_0} &= 1 + \frac{3}{4 k_1} \textrm{Re} \left[ \int_0^{+\infty} dk_\parallel \left[ \left(\frac{k_\parallel}{k_{z,1}}\right)r_s(k_\parallel) - \left( \frac{k_\parallel k_{z,1}}{k_1^2}\right) r_p(k_\parallel)\right]e^{2 i k_{z,1}z_0}\right] . \label{eq:dipSIpar}
\end{align}
For an out-of-plane dipole, the total emission rate is:
\begin{align}
    \frac{\Gamma(z_0)}{\Gamma_0} &=  1 + \frac{3}{2 k_1^3} \textrm{Re}\left[ \int_0^{+\infty} dk_\parallel \left( \frac{k_\parallel ^3}{k_{z,1}}\right) e^{2ik_{z,1} z_0} r_p(k_\parallel)\right] \label{eq:dipSIperp}
\end{align}
and we note that this dipole orientation only contains p-polarized components. \par

For the cavity structure, we consider a dipole at $z_0 = 0$ in medium $1$ with permittivity $\epsilon_1$, which is symmetrically located between two media of permittivity $\epsilon_2$ and where the distance between their two interfaces is given by $d$ (figure \ref{fig:schem}(b)). The techniques described above can easily be generalized here. We again write the total field as $\textbf{E}(\boldsymbol{r})=\textbf{E}_{free}(\boldsymbol{r})+\textbf{E}_{sc}(\boldsymbol{r})$. The scattered field can be obtained from the reflected field of a single interface, by summing its propagation and multiple reflection to all orders. At the position of the dipole, one finds:
\begin{align}
    \textbf{E}_{sc} (\boldsymbol{r_0}) &= \frac{1}{4 \pi^2} \int d^2 \boldsymbol{k_\parallel} \left( \boldsymbol{E}_{sc}^+ (\boldsymbol{k_\parallel}) + \boldsymbol{E}_{sc}^- (\boldsymbol{k_\parallel}) \right) e^{i k_{z, 1} d/2}. 
\end{align}
The scattered field components are given by
\begin{align}
    \boldsymbol{E}_{sc}^{\pm} (\boldsymbol{k_\parallel}) &= \frac{i \omega^2 \mu_0}{2} \left[ \left(\frac{r_p  e^{i k_{z,1} d/2}}{1 + r_p  e^{i k_{z,1} d}} \right)\textbf{M}_R^{(p, \pm)}(\boldsymbol{k_\parallel}) + \left( \frac{r_s  e^{i k_{z,1} d/2}}{1 - r_s e^{i k_{z,1} d}}\right) \textbf{M}_R^{(s)}(\boldsymbol{k_\parallel})\right] \cdot \textbf{p} \label{eq:dipcavpar}
\end{align}
for an in-plane dipole, while for an out-of-plane dipole, 
\begin{align}
    \boldsymbol{E}_{sc}^{\pm} (\boldsymbol{k_\parallel}) &= \frac{i \omega^2 \mu_0}{2} \left[ \left(\frac{r_p  e^{i k_{z,1} d/2}}{1 - r_p  e^{i k_{z,1} d}} \right)\textbf{M}_R^{(p, \pm)}(\boldsymbol{k_\parallel}) \right] \cdot \textbf{p}. \label{eq:dipcavperp}
\end{align}
\par
From equation (\ref{eq:dip}), the corresponding emission rate for the in-plane dipole in the cavity is:
\begin{align}
    \frac{\Gamma(d)}{\Gamma_0} = 1 + \frac{3}{2 k_1} \textrm{Re} \left[ \int_0^{+\infty} dk_\parallel \left[  \frac{k_\parallel}{k_{z,1}} \left( \frac{r_s}{1-r_s e^{ik_{z,1} d}}\right) - \frac{k_\parallel k_{z,1}}{k_1^2}\left( \frac{r_p}{1 + r_p e^{i k_{z,1} d}}\right) \right]e^{i k_{z,1} d}\right].
\end{align}
\par 
Similarly, for the out-of-plane dipole,
\begin{align}
    \frac{\Gamma(d)}{\Gamma_0}  &= 1 + \frac{3}{k_1^3} \textrm{Re} \left[ \int_0^{+\infty} dk_\parallel \left(\frac{k_\parallel ^3}{k_{z,1}}\right) e^{ik_{z,1}d} \left(\frac{r_p}{1-r_p e^{ik_{z,1}d}} \right)\right] .
\end{align}

\subsection{Plane-dipole Emission Rates \label{sec:planeSER}}
While the point-dipole model accurately describes a number of quantum optical emitters, it is not necessarily suitable to model an exciton with a delocalized center of mass coordinate, nor the possible dependence of emission on the center of mass temperature.  Defining $\textbf{Q}=(Q_x,Q_y)$ as the in-plane, center-of-mass momentum \textbf{Q} of the exciton, the emitted photon will necessarily have the same in-plane momentum in addition to an out-of-plane momentum component that we denote by $Q_z^2 = k_1^2 - \textbf{Q}^2$. In analogy to the point dipole case, the spontaneous emission can be calculated by modeling the extended exciton as a \textit{planar} dipole, with current density~[7, 19]
\begin{align}
    \textbf{J}(\boldsymbol{r}) &= \textbf{J}_0 \delta(z-z_0) e^{i\textbf{Q}\cdot \boldsymbol{\rho}} e^{-i\omega t}.
\end{align}
Here, $\boldsymbol{\rho}$ denotes the in-plane coordinate, while $z_0$ is the position of the planar dipole along $z$. The spontaneous emission rate of the exciton of momentum $\textbf{Q}$ is given by 
\begin{align}
    \Gamma (\textbf{Q}) = - \frac{\alpha}{|J_0|^2}\textrm{Re}\left[ \int d\boldsymbol{r} \textbf{J}^\dagger(\boldsymbol{r}) \cdot \textbf{E}(\boldsymbol{r})\right]. \label{eq:serQ}
\end{align}
Here, $\alpha$ is a geometry-independent coefficient, which only depends on the microscopic properties of the exciton. For a point dipole, a natural choice to remove this coefficient is made by normalizing the emission rate relative to a uniform medium, as in equation (\ref{eq:dip}). For an extended exciton, however, we will soon see that an analogous choice can be problematic, and we will present several complementary possibilities.

To illustrate how the emission of an exciton differs from a point dipole, we begin by recovering well-known results~[30] regarding radiative emission of an exciton in a uniform medium of permittivity $\epsilon_1$. In that case, we can exploit equations (\ref{eq:efieldgreen})-(\ref{eq:m}) to find the free~(and total) field, 

\begin{align}
    \textbf{E}(\boldsymbol{r})=\textbf{E}_{free}(\boldsymbol{r}) = -\frac{\omega \mu_0}{2} e^{i(\textbf{Q} \cdot \boldsymbol{\rho} + Q_z |z-z_0|)}e^{-i \omega t}\textbf{M}(\textbf{Q}) \cdot \boldsymbol{J_0}. \label{eq:plfield}
\end{align}
First, we can consider an exciton with a circular, in-plane transition, corresponding to $\boldsymbol{J_0} = J_0 \hat{\sigma}^{\pm} = \frac{J_0}{\sqrt{2}}(\hat{x} \pm i\hat{y})$ (where the sign difference corresponds to left-handed or right-handed polarization). In that case, a natural choice to normalize emission rates is relative to the emission rate of a stationary exciton in the uniform medium, which from equation (\ref{eq:serQ}) yields:
\begin{align}
    \frac{\Gamma_0(Q)}{\Gamma_0(\textbf{Q}=0)} &= \frac{1}{2}\left[ \frac{Q_z^2 + k_1^2}{Q_z k_1}\right] \label{eq:servac}
\end{align}
for $Q_z<k_1$, and $\Gamma_0(Q)=0$ otherwise. Here, the subscript in $\Gamma_0(Q)$ refers to the emission in a uniform medium. Notably, the above result shows that the exciton can only radiate when its momentum can match that of a propagating photon.

On the other hand, for an out-of-plane transition, $\boldsymbol{J_0} = J_0 \hat{z}$, one finds
\begin{align}
    \Gamma_0(Q) &= \frac{\alpha \omega \mu_0}{2 k_1} \left(\frac{Q^2}{k_1 Q_z}\right) \label{eq:perpvac}
\end{align}
for $Q_z<k_1$, and $\Gamma_0(Q)=0$ otherwise. Unlike the in-plane exciton, a stationary out-of-plane exciton is dark, $\Gamma_0(\textbf{Q}=0)=0$, so the previous normalization procedure cannot be applied here. One practical and experimentally meaningful way to normalize can be to consider a non-zero motional temperature for the momentum $\textbf{Q}$, and calculating the thermally averaged emission rate, which we will introduce later. \par

The calculation of the spontaneous emission rate of an exciton near a planar interface (figure \ref{fig:schem}(c)) proceeds in an analogous fashion to the point dipole.
It can readily be shown that for an in-plane, circularly polarized transition, the total emission rate is given by
\begin{align}
    \frac{\Gamma (Q, z_0)}{\Gamma_0(Q=0)} &= \frac{1}{2} \left( \frac{Q_{z,1}^2 + k_1^2}{Q_{z,1} k_1} + \textrm{Re}\left[ - \frac{Q_{z,1}}{k_1} r_p e^{2i Q_{z,1} |z_0|}\right] + \textrm{Re}\left[ \frac{k_1}{Q_{z,1}} r_s e^{2i Q_{z,1} |z_0|}\right]\right), \label{eq:SIserQpar}
\end{align}
where we have normalized the emission rate by that of a stationary exciton in the uniform medium. For the out-of-plane exciton, the total emission rate is:
\begin{align}
     \Gamma (Q, z_0) &= \frac{\alpha \omega \mu_0}{2 k_1} \left(  \frac{Q^2}{k_1 Q_{z,1}} + \textrm{Re}\left[ r_p \frac{Q^2}{k_1 Q_{z,1}} e^{2i Q_{z,1} |z_0|} \right] \right). \label{eq:SIserQperp}
\end{align}

\par
Likewise, in the cavity structure illustrated in figure \ref{fig:schem}d, the emission rate for an in-plane, circularly polarized transition is
\begin{align}
    \frac{\Gamma(Q, d)}{\Gamma_0(Q=0)} &= \frac{1}{2} \left[\frac{(Q_{z,1})^2 + k_1^2}{Q_{z,1} k_1} + \textrm{Re} \left[\left( \frac{k_1}{Q_{z,1}}\right) \frac{2 r_s e^{iQ_{z,1} d}}{1-r_s e^{i Q_{z,1} d}}\right] + \textrm{Re}\left[-\left( \frac{Q_{z,1}}{k_1}\right) \frac{2 r_p e^{iQ_{z,1} d}}{1+r_p e^{i Q_{z,1} d}} \right]\right], \label{eq:CAVserQpar}
\end{align}
while for the out-of-plane transition,
\begin{align}
    \Gamma (Q, d) &= \frac{\alpha \omega \mu_0}{2 k_1} \left(\frac{Q^2}{k_1 Q_{z,1}} + \textrm{Re} \left[ \left(\frac{Q^2}{k_1 Q_{z,1}}\right) \frac{2 r_p e^{i Q_{z,1}d}}{1- r_p e^{i Q_{z,1}d}}\right] \right). \label{eq:CAVserQperp}
\end{align}
\par 

\section{Results: modified emission by silver interface and cavity \label{sec:dipres}}


\subsection{Point dipole}

With the preceeding formalism, we now consider the modified emission rates of point dipoles in the planar structures detailed in figure \ref{fig:schem}. In these calculations, we consider metal interfaces of silver. We take the permittivity of silver to be $\epsilon_s = - 27.397 + 0.896i$ [31], which is accurate at the wavelength of the neutral exciton transition in MoSe$_{2}$ ($\lambda = 755$ nm) [19]. For simplicity, we consider the dielectric medium that surrounds the exciton to be vacuum ($\epsilon_1 = 1$). \par  

\begin{figure}[h!]
\centering
\includegraphics[width = 0.5\textwidth]{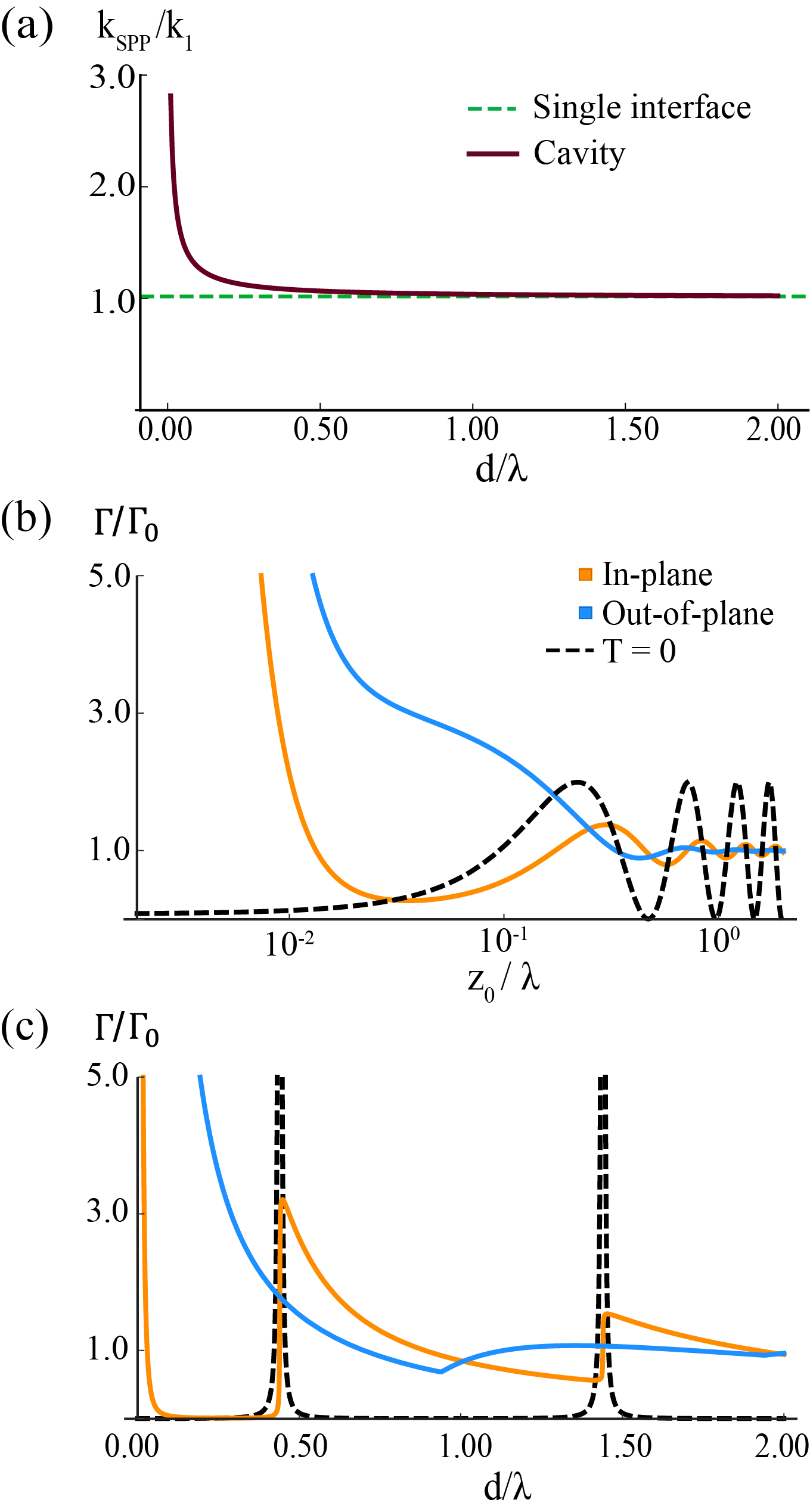}
\caption{(a) Plasmon dispersion relation $k_{spp}/k_1$ for a symmetric silver-vacuum cavity~(solid curve), as a function of mirror separation $d$. Note that as the distance between the interfaces increases, the SPP wavevector approaches the value for a single interface ($k_{spp} \approx 1.019 k_1$, dashed line). (b) Normalized $\Gamma/\Gamma_0$ emission rate for in-plane and out-of-plane point dipoles as a function of distance $z_0$ from a single silver interface. The emission rate for an extended exciton with an in-plane dipole transition at $T=0$ is also included. (c) Normalized emission rate for in-plane and out-of-plane point dipoles in a symmetric silver cavity versus mirror separation $d$, as well as the emission rate for an in-plane extended exciton at $T=0$.}
\label{fig:dipresults}
\end{figure}

In the single-interface case, the reflection coefficient $r_p(k_\parallel)$ (equation (\ref{eq:rsp})) contains a pole if $\epsilon_s$ is negative, corresponding to a surface plasmon polariton (SPP) mode with in-plane wavevector component~[32]
\begin{align}
    k_{spp} &= \sqrt{\frac{\epsilon_1 \epsilon_s (\omega)}{\epsilon_1 + \epsilon_s (\omega)}} \approx 1.019 k_1
\end{align}
for the silver parameters given above. For the symmetric silver cavity, the SPP wavevector is given by the solution to
\begin{align}
    1 -  r_p^2(k_{spp}) & e^{2 i k_{z,1} d} = 0.
\end{align}
We plot the solutions for $k_{spp}$ as a function of the cavity separation $d$ in figure \ref{fig:dipresults}(a). \par
In figure \ref{fig:dipresults}(b), we plot the modified emission rates for the in-plane and out-of-plane point dipoles at a single silver interface (equations (\ref{eq:dipSIpar}) and (\ref{eq:dipSIperp})), as a function of distance $z_0$. At large distances, the emission rate behavior is oscillatory, resulting from interference between the emitted and reflected fields. Moreover, the short distance~($z_0\rightarrow 0$) behavior scales as $1/z_0^3$ and is proportional to the silver loss~(Im~$\left[\epsilon_s\right]$), reflecting the energy loss produced by Ohmic dissipation of the induced currents in the silver. This contribution arises from the large $k_{\parallel}>k_1$ tails of the integrands in equations~(\ref{eq:dipSIpar}) and (\ref{eq:dipSIperp}). At intermediate distances, the out-of-plane dipole emits significantly into the SPP modes, due to the purely p-polarized nature of both the dipole radiation and the SPP modes. Assuming that $\epsilon_s$ is real and evaluating the pole contribution to equation (\ref{eq:dipSIperp}) gives:
\begin{align}
    \frac{\Gamma_{spp}}{\Gamma_0} &\approx 1 - \frac{3 \pi k_{spp}^3}{\epsilon_1^{3/2} k_1^2 \sqrt{k_{spp}^2 - k_1^2}} \textrm{Re} \left[ \frac{\epsilon_s^{3/2}}{\sqrt{1+\epsilon_s} (1 + \epsilon_s^2)}\right] e^{-2 |k_{z, spp}| z_0} . \label{eq:sppSI}
\end{align}

We plot the modified emission rate versus $d$ for the cavity in figure \ref{fig:dipresults}(c). In this case, the in-plane dipole exhibits regions of enhancement and suppression with a saw-tooth behavior, in agreement with the well-known results for electric dipole transitions [22]. Coupling to SPP modes is prohibited for the in-plane dipole for this symmetric cavity geometry. For the out-of-plane orientation, a large enhancement of emission is observed at short distances $d/\lambda \lesssim 0.5$, which is attributable to the SPP modes. The approximate SPP emission rate can be calculated in the near-field~($d\rightarrow 0$) limit, where the SPP wavevector can be approximated as:
\begin{align}
    k_{spp} \approx \ln{\left[\frac{\epsilon_s - \epsilon_1}{\epsilon_s + \epsilon_1}\right]} \frac{1}{d} .\label{eq:kspp_cavity}
\end{align}
The corresponding tight field confinement and small mode volume at small $d$ give rise to a significantly enhanced decay rate into SPPs:
\begin{align}
    \frac{\Gamma_{spp}}{\Gamma_0} &\approx \frac{3 \pi}{(k_1 d)^3} \textrm{Re} \left[ \ln^2{\left[ \frac{\epsilon_s - \epsilon_1}{\epsilon_s + \epsilon_1}\right]}\right].
\end{align}
Similar to the single-interface analysis, at short distances $d\rightarrow 0$ both dipole orientations also experience an emission rate that scales like $(\textrm{Im}\,\epsilon_s)/d^3$ due to Ohmic dissipation. However, this contribution always remains smaller than the SPP emission for the out-of-plane dipole, which also has a $1/d^3$ scaling.


\clearpage 
\subsection{Momentum-resolved emission rate of exciton}



\begin{figure}[h!]
\centering
\includegraphics[width = 0.9\textwidth]{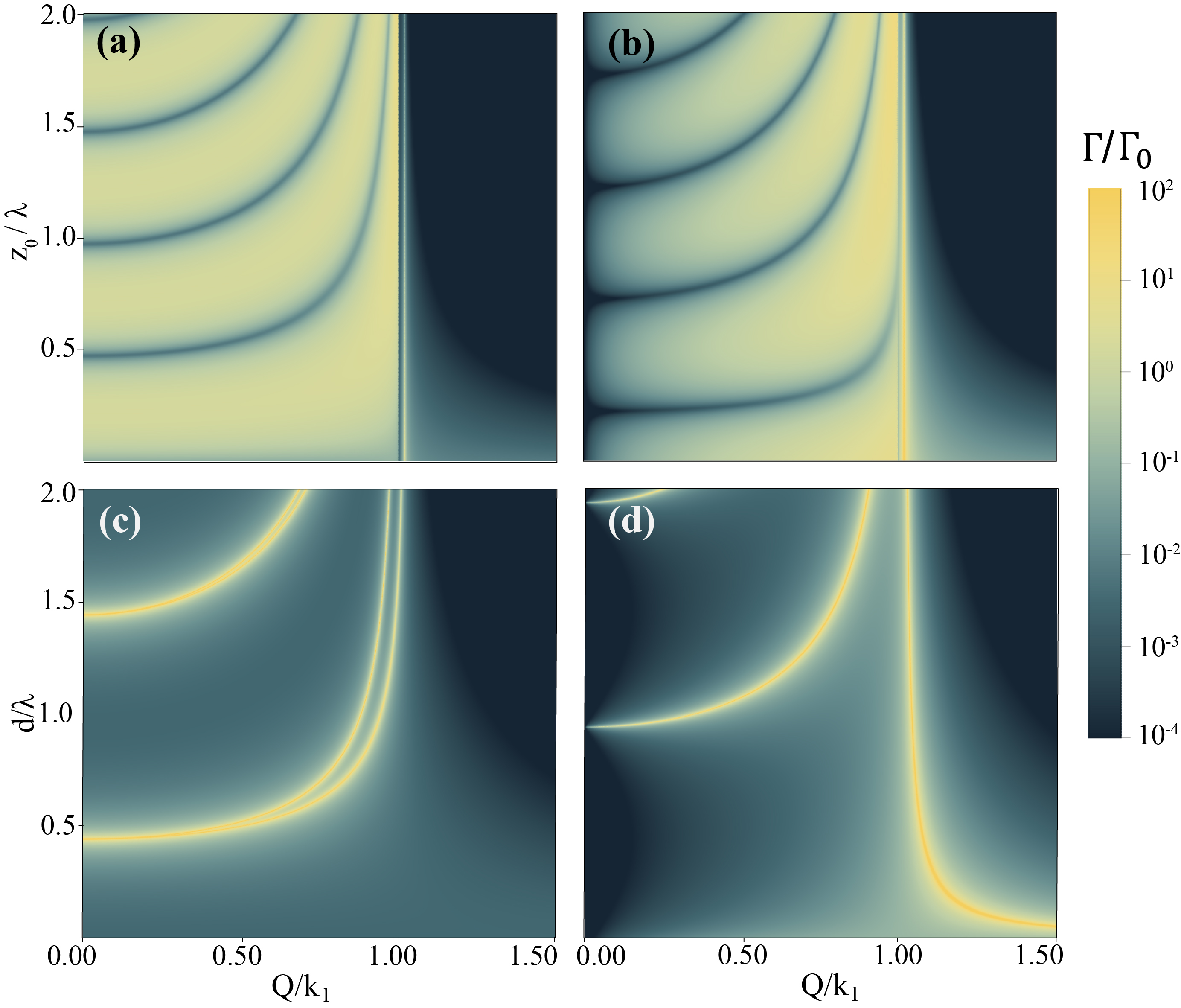}
\caption{Normalized emission rates as a function of center-of-mass momentum $Q/k_1$ and distance between exciton and interface $z_0$ (or mirror separation $d$ for cavity case). \textbf{(a)} $\Gamma(Q,z_0)/\Gamma_0(Q=0)$ for a single interface, and in-plane circularly polarized planar dipole. \textbf{(b)} $\Gamma(Q,z_0)/\Gamma_0(Q=k_1/\sqrt{2})$ for a single interface, and out-of-plane polarized planar dipole. \textbf{(c)} $\Gamma(Q,d)/\Gamma_0(Q=0)$ for a cavity and in-plane circularly polarized planar dipole. \textbf{(d)} $\Gamma(Q,d)/\Gamma_0(Q=k_1/\sqrt{2})$ for a cavity and out-of-plane polarized planar dipole.}
\label{fig:excres}
\end{figure}

In this section, we present results for the momentum-resolved emission rate of an extended exciton for the single interface and cavity structures. Figure \ref{fig:excres}(a) shows the normalized emission rate $\Gamma(Q,z_0)/\Gamma_0(Q=0)$ for a single interface as a function of the center-of-mass momentum $Q$ and the distance $z_0$ from the surface for an in-plane transition as given by equation (\ref{eq:SIserQpar}). In figure \ref{fig:excres}(b), we plot an analogous quantity $\Gamma(Q,z_0)/\Gamma_0(Q=k_1/\sqrt{2})$ for the out-of-plane case, equation~(\ref{eq:SIserQperp}). Here, we have chosen the (arbitrary) normalization of an exciton in a uniform medium with momentum $Q=k_1/\sqrt{2}$, as $Q=0$ yields a zero emission rate. In both the in-plane and out-of-plane cases, the emission rate for a fixed momentum in the radiative region ($Q < k_1$) is oscillatory as the exciton position $z_0$ is varied, much like the oscillations seen in the point dipole case. We note, however, that for the extended exciton, the visibility of the oscillations does not decrease with increasing $z_0$. In the non-radiative region ($Q > k_1$), one observes an emission into SPP's when $Q=k_{spp}$, and lossy emission for other $Q$, provided that the exciton is close enough to couple evanescently, $z_0\lesssim 1/|Q_z|$.\par

One particular limit is the case of zero motional temperature, $Q=0$. In figure \ref{fig:dipresults}(b), we compare the normalized emission $\Gamma(Q=0,z_0)/\Gamma_0(Q=0)$ for an in-plane transition of an extended exciton at zero temperature with that $\Gamma(z_0)/\Gamma_0$ of a point dipole. The extended exciton curve is notably different in its undamped oscillations with $z_0$, and the absence of a near-field non-radiative emission for small $z_0$. In the out-of-plane case, the extended exciton has a zero-temperature emission rate $\Gamma(Q=0,z_0)=0$, unlike the point dipole case.

In figures \ref{fig:excres}(c) and (d), we repeat the calculations of the momentum and distance dependent emission rates, $\Gamma(Q,d)/\Gamma_0(Q=0)$ and $\Gamma(Q,d)/\Gamma_0(Q=k_1/\sqrt{2})$, for in-plane and out-of-plane transitions, respectively, this time for the cavity geometry. Now in the radiative region $Q<k_1$, the emission is sharply enhanced for cavity separation distances $d$ that yield resonances at the exciton emission frequency. These resonance conditions are determined by the poles of the terms in equations (\ref{eq:CAVserQpar}) and (\ref{eq:CAVserQperp}). In the case of a nearly perfect conductor $\epsilon_s\rightarrow-\infty$, the resonance condition at normal incidence $Q=0$ would occur at $d/\lambda = m + 1/2$ for non-negative integer $m$. The splitting of the resonance condition for $Q\neq 0$, observed in the in-plane case, arises due to the slight phase difference between the s- and p-reflection coefficients for the finite permittivity of silver. Similar to the single interface, the emission rate is periodic in $d$, but now with a Purcell-enhanced maximum on resonance that scales with the cavity finesse, $\sim 1/(1-|r|^2)$. For $Q>k_1$, the out-of-plane transition also exhibits a sharp feature following the SPP dispersion relation given in equation (\ref{eq:kspp_cavity}), $Q\approx k_{spp}$, while the in-plane transition does not couple to SPP's for the symmetric geometry considered here. 


\clearpage 
\section{Temperature dependent emission rates \label{sec:temp}}

Next, we consider how the temperature dependence of the exciton momentum distribution impacts our results for emission rates. The distribution can be governed by the interplay of various complex microscopic mechanisms, including radiative and non-radiative recombination rates and thermalization due to scattering with phonons and other excitons [33-36]. Additionally, excitons are bosonic quasiparticles that can form (quasi-)condensed phases, significantly modifying the momentum distribution. Here, we calculate the resulting spontaneous emission rate of a thermal equilibrium distribution of a non-interacting bosonic gas with a fixed temperature $T$, assuming that the momentum distribution always relaxes on time scales that can be considered instantaneous compared to the emission time. In TMDCs, thermalization occurs on a timescale of picoseconds - therefore, the assumption of thermalization is valid for excitons with lifetimes significantly exceeding this timescale [33].

Under this assumption, the Bose-Einstein~(BE) distribution for the momentum is given by:
\begin{align}
    f_{BE}(\epsilon) &= \frac{1}{e^{(\epsilon(Q) - \mu)/k_B T} - 1}
\end{align}
where $\epsilon(Q) = \frac{(\hbar Q)^2}{2 m}$, $m$ is the exciton mass, and $\mu$ is the chemical potential, related to the density of particles $n$ by
\begin{align}
    \mu &= k_B T \ln{\left( 1 - e^{-n \lambda_{dB}^2}\right)}.
\end{align}
The transition between quantum and classical statistics is governed by the dimensionless phase space density $n \lambda_{dB}^2$, where $\lambda_{dB} = \frac{h}{\sqrt{2 \pi m k_B T}}$ is the thermal de Broglie wavelength. When $n \lambda_{dB}^2 \gtrsim 1$, the quantum statistics become significant. Although the case of 2D does not produce a true condensate with a macroscopically occupied ground state [37], we will later see that the emission nonetheless exhibits a modified behavior similar to an ideal condensate or a $T=0$ gas. On the other hand, for $n \lambda_{dB}^2 \ll 1$, the BE distribution approaches the classical Maxwell-Boltzmann~(MB) formula with a~(per-particle) distribution of,
\begin{align}
    f_{MB}(\epsilon) = \frac{\hbar^2 }{2 \pi m k_B T} e^{- \epsilon(Q)/k_B T}. \label{eq:MBdist}
\end{align}

Under the above assumptions, the average spontaneous emission rate $\langle \Gamma \rangle$ at a given density and temperature is given by
\begin{align}
    \langle \Gamma\rangle &= \frac{ \int \Gamma(\textbf{Q}) f_{BE}(\textbf{Q}) d^2 \textbf{Q}}{\int f_{BE}(\textbf{Q}) d^2\textbf{Q}}. \label{eq:beSER}
\end{align}
One convenient and experimentally meaningful way to normalize this quantity is to divide it by the average emission rate at the same temperature and particle density $n$, but in free space, which we denote by $\langle \Gamma_0 \rangle$. Note that for non-zero temperature, $\langle \Gamma_0 \rangle$ is non-zero for both in-plane and out-of-plane dipoles, and thus this normalization can be equally applied to each case. 


\section{Results: temperature dependence at a single interface \label{sec:SI}}

\subsection{In-plane polarization}\label{subsec:inplane}

\begin{figure}[ht!]
\centering
\includegraphics[width = \textwidth]{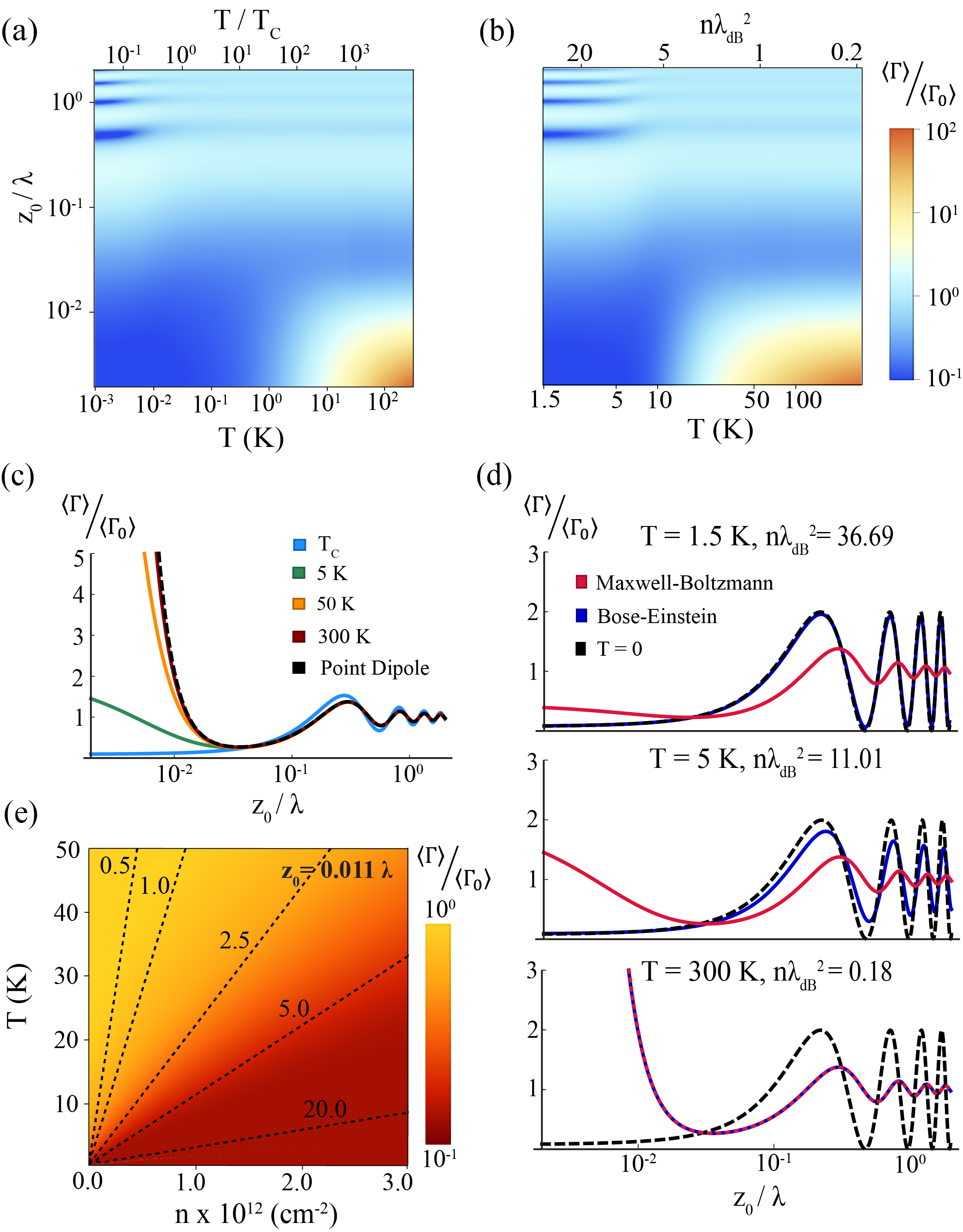}
\caption{(Caption on next page.)}
\label{fig:SIpar}
\end{figure}

Here, we study the radiative emission of thermalized in-plane polarized excitons near a single metal interface using the same material parameters as in section \ref{sec:dipres}. We take the exciton mass to be $m = m_e + m_h$ where $m_e = 0.49 m_0$ and $m_h = 0.52 m_0$ for MoSe${}_2$ are obtained from DFT calculations ($m_0$ is the electron mass) [38, 39]. For the MB distribution (equation (\ref{eq:MBdist})), we use the $Q$-dependent emission rates given in equations (\ref{eq:servac}) and (\ref{eq:SIserQpar}) to calculate the normalized, average emission rate, $\langle \Gamma\rangle/\langle \Gamma_0 \rangle$, for the in-plane, cicularly polarized plane dipole at non-zero temperatures. We plot the thermally averaged emission rate as a function of temperature and distance $z_0$, in figure \ref{fig:SIpar}(a) and provide the corresponding value of $T / T_c$ on the upper axis. The characteristic energy scale defined by $k_B T_c = (\hbar k_1)^2/2m$ corresponds to when the exciton has the same momentum as the radiative photon~($T_c \sim 30$ mK for our parameters). We find that for $z_0 > 0.5 \lambda$, the emission rates for behave similarly with temperature, oscillating interferometrically with a visibility that increases for $T\lesssim T_c$. However, at small distances, when the 2D TMDC is placed close to the metal interface ($z_0 < 0.03 \lambda$), the differences in their temperature dependence become apparent. This is evident in figure \ref{fig:SIpar}(c), where we plot the emission rate as a function of $z_0$ for various temperatures. Interestingly, as the temperature of the system increases, the functional form of $\langle \Gamma \rangle/\langle \Gamma_0\rangle$ approaches that of a point dipole, $\Gamma(z_0)/\Gamma_0$, as given by Eq.~(\ref{eq:dipSIpar}) and plotted with the dashed black curve. At sufficiently low temperatures ($T \sim T_{c}$), the exciton momentum distribution contains a negligible contribution from non-radiative components $Q>k_1$, which results in a significant quenching of non-radiative emission at short distances, as compared to the high temperature and point-dipole cases where non-radiative emission is dominant. \par 

We now calculate the average emission rate associated with a BE distribution, using Eqs.~(\ref{eq:servac}),~(\ref{eq:SIserQpar}), and~(\ref{eq:beSER}). Here, we choose a fixed total density of $n = 10^{12}$ cm$^{-2}$ typical of interlayer exciton systems [4] and use the same exciton mass $m$ as before. In Figure \ref{fig:SIpar}(b), we plot the emission rate $\langle \Gamma\rangle/\langle \Gamma_0 \rangle$ as a function of distance $z_0$ and temperature $T$. For convenience, we also indicate the corresponding value of $n\lambda_{dB}^2$ on the upper x-axis. In particular, one sees that a qualitative change in behavior occurs around $n\lambda_{dB}^2\sim 1$, with values $n\lambda_{dB}^2\lesssim 1$ approaching the MB emission properties discussed earlier. On the other hand, for values $n\lambda_{dB}^2\gtrsim 1$, bosonic enhancement of low-energy~($Q\sim 0$) states yields more prominent oscillations in the emission at large distances $z_0$, reminiscent of that of a planar dipole with center-of-mass momentum $Q=0$~(see figure~\ref{fig:excres}(a)). This similarity can be better seen in figure~\ref{fig:SIpar}(d), where we plot $\langle \Gamma\rangle/\langle \Gamma_0 \rangle$ as a function of distance, for several different temperatures. In particular, we compare the results obtained from the MB and BE distributions~(red and blue curves, respectively), with the emission calculated at $T=0$~(black curve). It is seen that for $n\lambda_{dB}^2\gg 1$, the BE curve essentially approaches the $T=0$ result, even though the temperature $T\gg T_c$ is still much larger than that to narrow the single-particle MB momentum distribution to only radiative components. Conversely, for $n\lambda_{dB}^2 \ll 1$, the emission curves for the MB and BE distributions coincide. Finally, in Figure~\ref{fig:SIpar}(e), we plot the emission rate $\langle \Gamma\rangle/\langle \Gamma_0 \rangle$ obtained by the BE distribution, as a function of temperature $T$ and total density $n$, for a fixed distance of $z_0 = 0.011 \lambda$ nm. The dashed contours indicate the values of $n\lambda_{dB}^2$. The transition around $n\lambda_{dB}^2\sim 1$ from the MB result to essentially the $T=0$ result~(for $n\lambda_{dB}^2 \gg 1$) is also evident here.

\addtocounter{figure}{-1}
\begin{figure}[t]
    \caption{(a, b) Temperature (MB and BE distributions respectively) and distance dependence of the normalized in-plane exciton emission rate, $\langle \Gamma \rangle/\langle \Gamma_0 \rangle$ near a single interface. At the top of (a), we provide the temperature normalized by $T_c$, while at the top of (b), we provide the temperature in terms of the product of particle density and de Broglie wavelength, $n\lambda_{dB}^2$. (c) Temperature linecuts comparing the plane-dipole emission rate (MB distribution) at various temperatures to the point-dipole model. (d) Comparison of emission rates using the MB and BE momentum distributions at various temperatures. (e) Emission rate with varying total densities $n$ and temperature $T$ at a fixed distance, $z_0=0.011 \lambda$ (colorbar uses a logarithmic scale). The contour lines denote the values of $n\lambda_{dB}^2$.}
\end{figure}
\clearpage 
\subsection{Out-of-plane polarization}
\begin{figure}[h!]
\centering
\includegraphics[width = \textwidth]{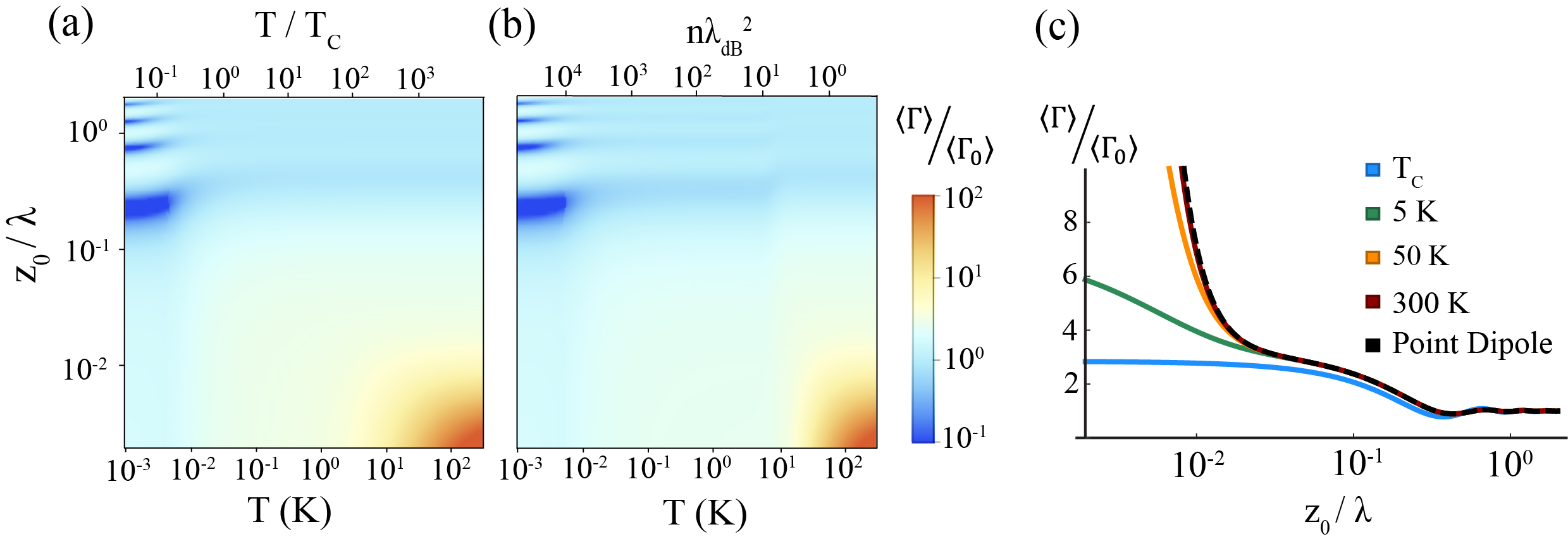}
\caption{(a,b) Temperature (MB and BE distributions respectively) and distance dependence of the normalized out-of-plane exciton emission rate, $\langle \Gamma \rangle/\langle \Gamma_0 \rangle$ near a single interface. (c) Temperature linecuts comparing the plane-dipole emission rate (MB distribution) at various temperatures to the point-dipole model.}
\label{fig:SIperp}
\end{figure}

We can repeat the calculations of the previous subsection, but now for an out-of-plane transition. In particular, in Figure~\ref{fig:SIperp}(a), we plot the normalized average emission rate $\langle \Gamma\rangle / \langle \Gamma_0 \rangle$ as a function of distance and temperature, assuming a MB distribution, while in Figure~\ref{fig:SIperp}(c), we provide line cuts showing the distance dependence for several specific temperatures. In Figure~\ref{fig:SIperp}(c), we also plot the normalized emission rate $\Gamma(z_0)/\Gamma_0$ for a point dipole~(dashed black curve). Just as in the case of an in-plane transition, the out-of-plane transition results converge to the point dipole one at sufficiently large temperatures and distances. Also as before, for small distances and low temperature, suppression of non-radiative emission is responsible for the divergence of the planar and point dipole results from one another. \par 
In Figure~\ref{fig:SIperp}(b), we plot the average emission rate versus distance and temperature for the BE distribution, taking the same total density $n = 10^{12}$ cm$^{-2}$ as in Sec.~\ref{subsec:inplane}. Again, one observes a qualitative transition in behavior around $n\lambda_{dB}^2 \sim 1$. 

\section{Results: temperature dependence in a metal cavity \label{sec:CAV}}

\subsection{In-plane polarization}

\begin{figure}[h!]
\centering
\includegraphics[width = 0.95\textwidth]{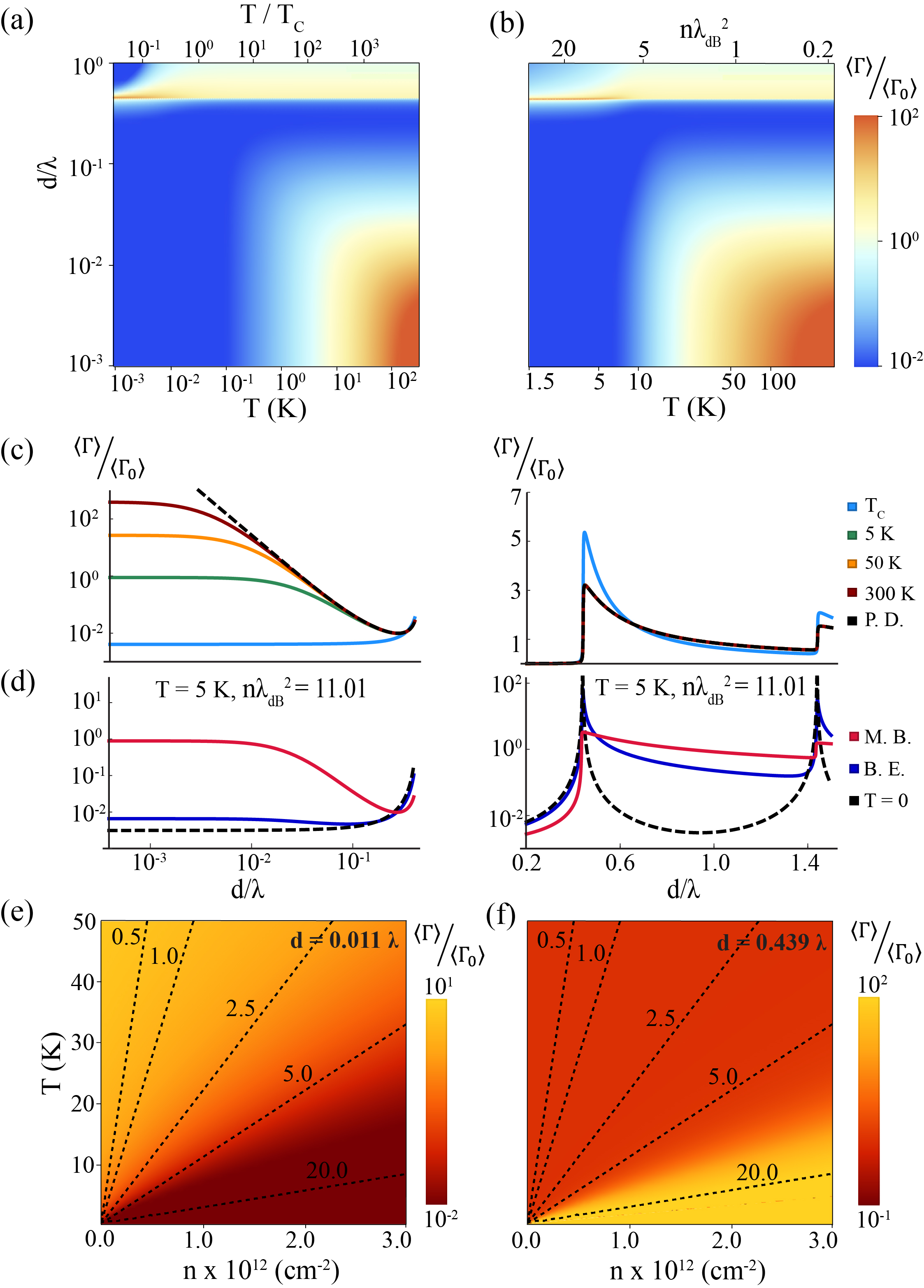}
\caption{(Caption on next page)} 
\label{fig:CAVpar}
\end{figure}

We now use Equations~(\ref{eq:servac}),~(\ref{eq:CAVserQpar}), and~(\ref{eq:beSER}) to calculate the normalized average emission rate for the cavity configuration. In Figures~\ref{fig:CAVpar}(a-b), we plot the temperature and distance~($d$) dependent rates for the MB and BE distributions, respectively. In the latter case, we again take a total density $n = 10^{12}$ cm$^{-2}$. In Figure~\ref{fig:CAVpar}(c), we plot linecuts of the MB results at select temperatures at small distances (left), and at larger distances (right), where the latter shows the strong sawtooth shaped Purcell enhancement associated with the cavity.   Similar to the single-interface case, we find that as the temperature increases, the plane-dipole model begins to behave like the point dipole model~(dashed black curves of Figure~\ref{fig:CAVpar}(c)) as the high-Q states that contribute to non-radiative emission become populated.  For $T \geq 5$K, we find that the minimum emission rate is $\langle \Gamma \rangle/\langle\Gamma_{0}\rangle \approx 0.01$ for a cavity separation of $d \sim 0.27 \lambda$ and is sharply enhanced for smaller separations, $d < 0.27 \lambda$, due to non-radiative decay. In comparison, this non-radiative decay becomes negligible at short distances once $T\sim T_c$. In the Purcell-enhanced region, it can be seen that low temperatures $T\lesssim T_c$ can result in a greater level of spontaneous emission enhancement, as one would expect as the $Q$ distribution narrows toward $Q=0$ (compare with Fig.~\ref{fig:excres}(c) at $Q=0$). \par 

For BE distributions with $n\lambda_{dB}^2\gtrsim 1$, bosonic enhancement allows for the normalized emission rate to come close to the $T=0$ case, similar to what was observed at a single interface. This is evident in Figure \ref{fig:CAVpar}(d), where we plot the distance-dependent emission rates at a fixed temperature of $T = 5$K ($n\lambda_{dB}^2 = 11.01$), using the MB and BE distributions~(red and blue, respectively), and also the $T=0$ result~(black dashed). Notably, for example, a maximum Purcell enhancement of $\langle \Gamma \rangle/\langle \Gamma_0 \rangle\sim 10^2$, as allowed for $T=0$, is also observed for the BE distribution.  In figures~\ref{fig:CAVpar}(e) and~\ref{fig:CAVpar}(f), we now consider the emission rate versus temperature and total density, at a close distance of $d=0.011\lambda$ and large distance of $d=0.439\lambda$, respectively. The transition from MB to BE behavior around $n\lambda_{dB}^2\sim 1$ is evident. Notably, at the distance of $d=0.439\lambda$, Purcell enhancements on the order of $\sim 10^2$ are observed when $n\lambda_{dB}^2\gg 1$. \par 
An important caveat to these discussions of orders of magnitude of enhancement is the possibility of strong coupling between the excitons and the metal cavity, in which case the calculations are no longer valid. For a metal cavity with $\sim 5$ THz linewidth, interlayer excitons with emission rates in the $10-100$MHz regime would remain weakly coupled even with the $\sim 100$ times Purcell enhancement shown in figure \ref{fig:CAVpar}(d) [3]. However, for individual monolayers where excitons can have THz radiative decay rates, the system would enter the strong coupling regime and the calculations are no longer valid [39].

\addtocounter{figure}{-1}
\begin{figure}[t]
\caption{(a,b) Temperature (MB and BE distributions respectively) and distance dependence of the normalized in-plane exciton emission rate, $\langle \Gamma \rangle/\langle \Gamma_0 \rangle$ in a symmetric silver cavity. At the top of (a), we provide the temperature normalized by $T_c$, while at the top of (b), we provide the temperature in terms of the product of particle density and de Broglie wavelength, $n\lambda_{dB}^2$. (c) Temperature linecuts comparing the plane-dipole emission rate (MB distribution) at various temperatures to the point-dipole model. (d) Comparison of emission rates using the MB and BE momentum distributions at $T=5$~K, and the comparison to the $T=0$ result. The left and right panels show the behavior for small and large cavity separations $d$, respectively. (e-f) BE emission rate versus temperature and particle density, for a small and large cavity separation, $d=0.011\lambda$ and $d=0.439\lambda$, respectively (colorbar is on a logarithmic scale). The contour lines indicate values of $n \lambda_{dB}^2$.}
\end{figure}

\clearpage 

\subsection{Out-of-plane polarization}

\begin{figure}[h!]
    \centering
    \includegraphics[width = \textwidth]{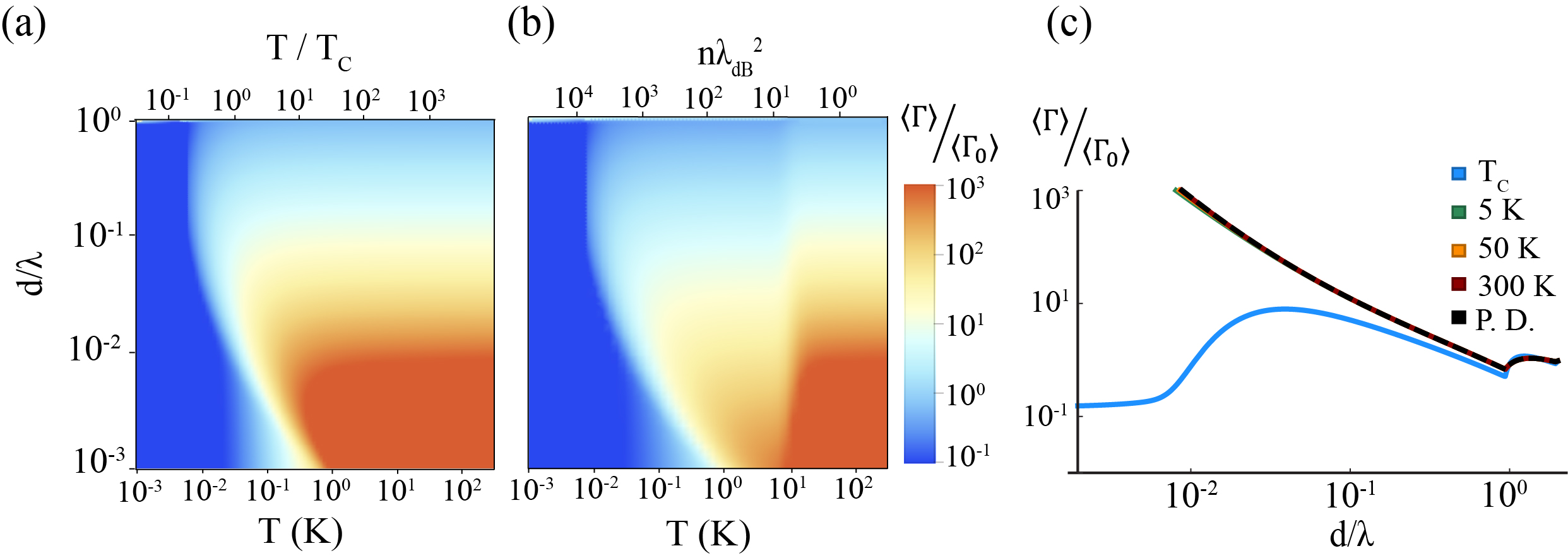}
    \caption{(a,b) Temperature (MB and BE distributions respectively) and distance dependence of the normalized out-of-plane exciton emission rate, $\langle \Gamma \rangle/\langle \Gamma_0 \rangle$ in a symmetric silver cavity. (c) Temperature linecuts comparing the plane-dipole emission rate (MB distribution) at various temperatures to the point-dipole model.}
    \label{fig:CAVperp}
\end{figure}

We now consider an out-of-plane transition in the cavity structure. Figure \ref{fig:CAVperp}(a) plots the emission rate dependence on temperature and cavity separation $d$ for the MB distribution. At low temperatures and small separations $d$, the large non-radiative emission present at high temperatures becomes significantly suppressed. Figure \ref{fig:CAVperp}(b) shows the emission rate calculated using the BE distribution. As before, a large phase space density allows the behavior associated with the MB distribution and low temperatures $T<T_c$ to be observed at much higher temperatures. Figure \ref{fig:CAVperp}(c) plots linecuts of the emission rate for the MB distribution at select temperatures. We find that at $T=T_c$, the non-radiative emission is suppressed by several orders of magnitude compared to higher temperatures and does not monotonically grow as $d\rightarrow 0$, unlike the point dipole case.  

\clearpage 
\section{Conclusion}\label{sec:conclusion}

We have elucidated the remarkably different emission behavior that extended excitons can have in common geometries such as a single metallic interface or a metallic cavity, as compared to the better known behavior of a point-like quantum emitter. We have shown how these differences are exaggerated when considering a Bose-Einstein distribution of the exciton center-of-mass momentum at high phase space densities $n\lambda_{dB}^2\gtrsim 1$. These insights may enable novel opportunities in optoelectronic systems and devices based on manipulating TMDC exciton emission.

\acknowledgements
DEC acknowledges support from the European Union’s Horizon 2020 research and innovation programme, under European Research Council grant agreement No 101002107 (NEWSPIN); the Government of Spain (Europa Excelencia program EUR2020-112155, Severo Ochoa program CEX2019-000910-S, and MCIN Plan Nacional Grant PGC2018-096844-B-I00); Generalitat de Catalunya through the CERCA program, AGAUR Project No. 2017-SGR-1334, Fundaci{\'o} Privada Cellex, and Fundaci{\'o} Mir-Puig. AB acknowledges support from the NSF Graduate Research Fellowship under grant no. DGE-1746045. GHC acknowledges support from the College Summer Undergraduate Research grant program and the Jeff Metcalf PME Fellowship at the University of Chicago.

\section{References}

\noindent [1] Mak K, Shan J 2016 \textit{Nature Photon} \textbf{10} 216-226. 
\newline 
[2] High A, Novitskaya E, Butov L, Hanson M, and Gossard A 2008 \textit{Science} \textbf{321} 5886.
\newline
[3] Jauregui L, \textit{et al}. 2019 \textit{Science} \textbf{366} 6467. 
\newline 
[4] Fogler M, Butov L, and Novoselov K 2014 \textit{Nat. Commun.} \textbf{5} 4555. 
\newline 
[5] Wang Z, Rhodes D, Watanabe K, Taniguchi T, Hone J, Shan J, and Mak K 2019 \textit{Nature} \textbf{574} 76-80. 
\newline 
[6] Zeytino\u{g}lu S, Roth C, Huber S, Atac Imamo\u{g}lu 2017 \textit{Phys. Rev. A} \textbf{96} 031801 (R).
\newline 
[7] Rogers C, Gray D, Bogdanowicz N, Taniguchi T, Watanabe K, and Mabuchi H 2020 \textit{Phys. Rev. Research} \textbf{2} 012029(R). 
\newline 
[8] Scuri G, \textit{et al.} 2018 \textit{Phys. Rev. Lett.} \textbf{120} 037402.
\newline
[9] Back P, Zeytino\u{g}lu S, Ijaz A, Kroner M, and Imamo\u{g}lu A 2018 \textit{Phys. Rev. Lett.} \textbf{120} 037401.
\newline 
[10] High A, Hammack L, \textit{et al}. 2012 \textit{Nature} \textbf{483} 584-588. 
\newline 
[11] Hijlkema M, Weber B, Specht H P, Webster S C, Kuhn A, and Rempe G 2007 \textit{Nature Phys} \textbf{3} 253-255.
\newline 
[12] Santori C, Fattal D, Vuckovic J, Solomon G S, and Yamamoto Y 2004 \textit{New J. Phys.} \textbf{6} 89.
\newline 
[13] Barros H G, Stute A, Northup T E, Russo C, Schmidt P O, and Blatt R 2009 \textit{New J. Phys.} \textbf{11} 103004.
\newline 
[14] Kuhn A, Hennrich M, and Rempe G 2002 \textit{Phys. Rev. Lett.} \textbf{89} 067901.
\newline 
[15] Kneipp K, Wang Y, Kneipp H, Perelman L T,  Itzkan I, Dasari R R, and Feld M S 1997 \textit{Phys. Rev. Lett.} \textbf{78} 1667.
\newline
[16] Etchegoin P G and Ru E C Le 2008 \textit{Phys. Chem. Chem. Phys.} \textbf{10} 6079-6089.
\newline 
[17] Jian Y, Chen S, Zheng W, Zheng B, and Pan A 2021 \textit{Ligh Sci Appl} \textbf{10} 72.
\newline
[18] Shi H, Yan R, Bertolazzi S, Brivio J, Gao B, Kis A, Jena D, Xing H G, and Huang L 2013 \textit{ACS Nano} \textbf{7} 1072-1080. 
\newline 
[19] Zhou Y \textit{et al.} 2017 \textit{Nature Nanotech} \textbf{12} 856-860.
\newline
[20] Chance R, Prock A, Silbey R 1978 Molecular Fluorescence and Energy Transfer Near Interfaces \textit{Advances in Chemical Physics} (\textbf{vol 37}) ed I Prigogine and S Rice (Wiley) p 1-65.
\newline 
[21] Drexhage K 1970 \textit{J. Lumin.} \textbf{1-2} 693-701.
\newline 
[22] Dutra S M and Knight P L 1996 \textit{Phys. Rev. A.} \textbf{53} 3587.
\newline 
[23] Noda S, Fujita M, and Asano T 2007 \textit{Nature Photon} \textbf{1} 449-458. 
\newline 
[24] Fan S, Villeneuve P R, Joannopoulos J D, and Schubert E F 1997 \textit{Phys. Rev. Lett.} \textbf{78} 3294. 
\newline 
[25] Fujita M, Takahashi S, Tanaka Y, Asano T and Noda S 2005 \textit{Science} \textbf{308} 1296-1298.
\newline 
[26] Agarwal G S 1975 \textit{Phys. Rev. A} \textbf{12} 1475. 
\newline 
[27] Novotny L and Hecht B 2006 \textit{Principles of Nano-Optics} (Cambridge: Cambridge University Press).
\newline 
[28] Agarwal G S 1998 \textit{J. Mod. Opt.} \textbf{45} 449.
\newline 
[29] Born M and Wolf E 1970 \textit{Principles of Optics} 6th edn (Oxford: Pergamon Press).
\newline 
[30] Wang H, Zhang C, Chan W, Manolatou C, Tiwari S, and Rana F 2016 \textit{Phys. Rev. B} \textbf{93}, 045407.
\newline 
[31] High A, Devlin R, Dibos A, Polking M, Wild D, Perczel J, de Leon N, Lukin M, and Park H 2015 \textit{Nature} \textbf{522} 192-196.
\newline 
[32] Heinz R 1988 \textit{Surface Plasmons on Smooth and Rough Surfaces and on Gratings} (\textit{Springer Tracts in Modern Physics} vol 111) (New York: Springer-Verlag).
\newline 
[33] Selig M, Bergh\"{a}user G, Richter M, Bratschitsch R, Knorr A, and Malic E 2018 \textit{2D Mater.} \textbf{5} 035017.
\newline 
[34] Umlauff M \textit{et al.} 1998 \textit{Phys. Rev. B} \textbf{57} 1390.
\newline 
[35] Ivanov A L, Littlewood P B, Haug H 1999 \textit{Phys. Rev. B} \textbf{59} 7. 
\newline 
[36] Mueller T, Malic E 2018 \textit{npj 2D Mater Appl} \textbf{2} 29.
\newline 
[37] Hohenberg P 1966 \textit{Phys. Rev} \textbf{158} 383.
\newline
[38] Wang G, Gerber I C, Bouet L, Lagarde D, Balocchi A, Vidal M, Amand T, Marie X, and Urbaszek B 2015 \textit{2D Mater.} \textbf{2} 045005.
\newline 
[39] Robert C \textit{et al.} 2016 \textit{Phys. Rev. B.} \textbf{93} 205423.
\end{document}